\documentclass[aps,prl,twocolumn,groupedaddress]{revtex4}%
\usepackage{graphicx}
\usepackage{delarray}
\usepackage{amsmath}
\usepackage{amsfonts}
\usepackage{amssymb}%
\begin{document}
\title{Remarks on quantum critical behavior in heavy fermions}
\author{Mucio A. \surname{Continentino} }
\email{mucio@if.uff.br}
\affiliation{Instituto de F\'{\i}sica - Universidade Federal Fluminense\\
Av. Litor\^anea s/n,  Niter\'oi, 24210-340, RJ - Brazil}
\date{\today}

\begin{abstract}
Generalized scaling relations and renormalization group results are used to discuss the phase diagrams of heavy
fermion systems. We consider the cases where these materials are driven to a magnetic quantum critical point
either by applying external pressure or a magnetic field.  The Ehrenfest equation relating the pressure derivative
of the critical temperature to the ratio of thermal expansion and specific heat close to a magnetic quantum
critical point (QCP) is analyzed from the scaling point of view. We consider different phase diagrams of
antiferromagnetic heavy fermions in an external uniform magnetic field and the implication of renormalization
group results for predicting their behavior.

\end{abstract}
\maketitle

Scaling theories are invaluable tools in the theory of quantum critical phenomena. Although these theories do not
predict the values of the critical exponents, they yield relations among them. Besides they show the relevant
physical quantities to be measured and  the exponents that can be extracted from their critical behavior. In the
case of heavy fermion systems, where there is no clear microscopic theory yet for their quantum criticality, the
scaling approach is specially useful \cite{mu}. In this Report, we use a generalized scaling theory of heavy
fermions \cite{mu} near an antiferromagnetic quantum instability to obtain some useful and general relations
involving measurable physical quantities and the exponents that control their critical behavior. We start
considering the case in which criticality is tuned by pressure and next the effects of an external magnetic field.

The Ehrenfest relation \cite{samara} when applied to magnetic systems close to a zero temperature magnetic
instability is a useful thermodynamic relation. This equation that relates  the pressure derivative of the line of
critical temperatures, $T_N$ to thermodynamic quantities at the quantum critical point (QCP) can be written as
\cite{samara},
\begin{equation}
\label{ehrenfest}
 \frac{dT_{N}}{dP}=VT\frac{\Delta\beta}{\Delta C}=V \frac{\Delta\beta}{\Delta
C/T}%
\end{equation}
where $\Delta\beta$ and $\Delta C$ are the differences in thermal expansion and specific heat in the two phases
(the critical part), respectively. The thermal expansion is defined by
\begin{equation}
\beta=\frac{1}{V}\frac{\partial V}{\partial T}=-\kappa_{T}\frac{\partial^{2}%
F}{\partial T\partial V}|_{N}%
\end{equation}
with the isothermal compressibility $\kappa_{T}=-\frac{1}{V}\frac{\partial V}{\partial P}$. The volume thermal
expansion, $\beta=2 \alpha_{a}+\alpha_{c}$ where $\alpha_{i}$ are the linear thermal expansion coefficients along
different axes $i$.

The expression for the critical line of finite temperature phase transitions close to the QCP is written as
\cite{mucio},
\[
T_{N}\propto|V-V_{c}|^{\psi}\propto|P-P_{c}|^{\psi}%
\]
which defines the shift exponent $\psi$. We have assumed that the critical temperature of the phase transition is
reduced  by changing the volume, for example, by applying pressure in the system. The quantities $V_{c}$ and
$P_{c}$ are the critical volume and pressure, respectively. From the equation above we obtain,
\begin{equation}
\label{deri}
\frac{dT_{N}}{dP}\propto|P-P_{c}|^{\psi-1}\propto T_{N}^{1-\frac{1}{\psi}}.%
\end{equation}
In the usual theory of quantum spin density wave transitions \cite{millis} the shift exponent can be expressed in
terms of the dynamic exponent $z$ and the dimensionality of the system $d$, for $d+z>4$, as   $\psi=z/(d+z-2)$, in
which case we get,
\begin{equation}
\frac{dT_{N}}{dP}\propto|P-P_{c}|^{\psi-1}\propto T_{N}^{\frac{2-d}{z}}%
\end{equation}

Equation \ref{deri} can be also obtained  using the Ehrenfest relation and the generalized scaling form of the
singular part of the free energy density which is given by \cite{mucio},
\begin{equation}
f\propto A(T)|V-V_{C}-uT^{1/\psi}|^{2-\tilde{\alpha}}%
\end{equation}
where
\[
A(T)=T^{\frac{{}^{\tilde{\alpha}-\alpha}}{\nu z}}%
\]
In these equations, the {\em tilde} exponents are associated with the finite temperature phase transition and the
{\em non-tilde} with the quantum critical point, since in general these transitions are in different universality
classes \cite{mucio}. For the coefficient of thermal expansion, we get
\begin{equation}
\Delta\beta(V=V_{c})\propto\frac{\partial^{2}f}{\partial T\partial
V}|_{V=V_{c}}\propto u^{1-\tilde{\alpha}}T^{\frac{{}^{\tilde{\alpha}-\alpha}%
}{\nu z}-1+\frac{{}^{1-\tilde{\alpha}}}{\psi}}%
\end{equation}
For the specific heat we get \cite{mucio},
\[
\Delta C/T \mid_{V=V_{c}}\propto\frac{\partial^{2}f}{\partial T^{2}}|_{V=V_{c}%
}\propto u^{2-\tilde{\alpha}}T^{\frac{{}^{\tilde{\alpha}-\alpha}}{\nu z}%
+\frac{^{2-\tilde{\alpha}}}{\psi}-2}%
\]
and finally we obtain for the ratio \cite{zhu,flouquet},
\begin{equation}
\label{ratio}
\frac{\Delta\beta}{\Delta C/T}|_{V=V_{c}}\propto\frac{u^{1-\tilde{\alpha}%
}T^{\frac{{}^{1-\tilde{\alpha}}}{\psi}-1}}{u^{2-\tilde{\alpha}}T^{\frac
{^{2-\tilde{\alpha}}}{\psi}-2}}=\frac{1}{u}T^{1-\frac{{}^{1}}{\psi}}%
\end{equation}
in agreement with Eq.~\ref{ehrenfest}. This is a general scaling result but in  the theory of quantum spin-density
wave transitions, for $d+z>4$, the quantity $u$ can be identified as the dangerous irrelevant quartic interaction
\cite{millis,mucio} of the action \cite{hertz}. Notice that working from the paramagnetic side of the quantum
critical point, the relevant exponent which governs the ratio above is the crossover exponent $\nu z$ \cite{zhu}.
This seems to imply the scaling relation $\psi = \nu z$ which identifies the shift  with the crossover exponent.
\cite{mu}. The spin density wave theory \cite{millis} makes clear the role of the dangerous irrelevant variable
$u$ in the violation of this scaling relation. Then the ratio given by Eq.~ref{ratio}, in a system which is driven
to a quantum critical point by pressure, {\em provides information on the shift exponent of the critical line}.
For a spin-density wave transition at $d+z>4$, it depends explicitly in the dangerous irrelevant variable $u$.

We now investigate the effects of  a magnetic field in the phase diagram of antiferromagnetic heavy fermion
systems \cite{mucio1,millis2}. We consider Ising-like systems with strong anisotropy and exclude from our
discussion the cases of phase diagrams with bicritical points. The main effect of the field that concerns us here
is that, when sufficiently strong, it can drive the transition temperature to zero. We discuss two different types
of phase diagrams.

\begin{figure}[ptb]
\includegraphics[width=6.5cm]{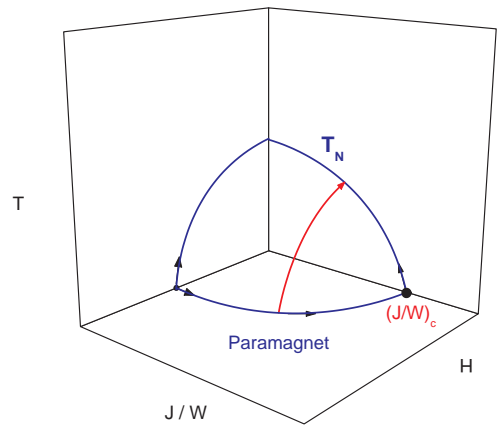}\caption{Possible phase diagram of an antiferromagnetic heavy
fermion in a magnetic field. The full lines are second order phase transitions and the arrows indicate the flow of the 
RG equations.  For $T \ne 0$ the field-driven transition has the same exponents of the $H=0$ N\'eel line. For $T=0$ this 
transition is controlled by the \protect$(J/W)_c$, \protect$H=0$ QCP.}
\label{fig1}%
\end{figure}

The most simple situation is that shown in Fig.~\ref{fig1}, where the magnetic field is a
completely irrelevant variable in the RG sense. From the expected flow of the RG equations \cite{dosSantos} shown in 
Fig.~\ref{fig1}, we find that if the system leaves the $AF$ phase by increasing the magnetic field at finite $T$,
the relevant exponents are those of the finite temperature, zero field N\'eel transition \cite{dosSantos}. On the
other hand if the system is driven, at zero temperature, to the paramagnetic phase by increasing the field, the
relevant exponents are those of the zero field quantum critical point at $(J/W)_c$ or $P_c$. Here $(J/W)$ is the
usual ratio between the Kondo lattice parameters \cite{mucio}. In this case, the critical line close to the PCQ
vanishes as $H_c \propto |(J/W) - (J/W)_c|^{1/2}$ with an analytic field dependence \cite{moss}.

\begin{figure}[ptb]
\includegraphics[width=6.5cm]{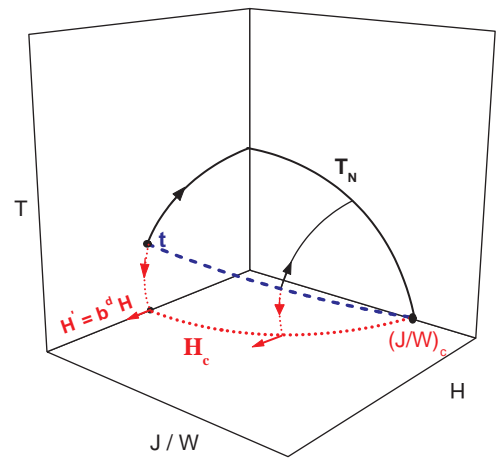}\caption{ The dotted  lines represent lines of first order transition and the 
continuous those of second order instabilities. The line of tricritical points \protect$t$
extends all the way to the \protect$H=0$ quantum critical point. The line \protect$H_c$ is a 
line of discontinuity fixed points \protect\cite{nienhuis} and it does not include the special multicritical point at 
\protect$(J/W)_c$, \protect$H=0$. }%
\label{fig2}%
\end{figure}

A more interesting case is that of Fig.~\ref{fig2}. Now there is a line of tricritical points ($t$) \cite{lawrie},
see Fig.~\ref{fig2}, separating the
lines of first and second order phase transitions that merges with the $H=0$ quantum critical point \cite{mucio1}. An 
essential feature in
this case is that {\em the magnetic field at the zero temperature line ($H_c$) containing the  end points of the lines 
of first order transitions scales as} $h^{\prime}=b^d h$, where $b$ is the scaling factor, $d$ the dimension of
the system and $h=H-H_c$ \cite{nienhuis,fisher}. The reason is that at this line the magnetic field has to reverse a 
large number of
spins of the order of the volume of the sample. Notice that this special line does not include the multicritical point 
at $H=0$.
The concept of finite size scaling \cite{fisher}, with the inverse of temperature $T$ playing the role of a cut-off 
length,
$L_{\tau} = T^{-1/z}$, and the scaling exponent of the magnetic field obtained before yields the  scaling form for the 
free energy at small but finite temperatures close to the line $H_c(J/W)$ or $H_c(P)$,
\begin{equation}
\label{free}
f \propto |h| F\left(\frac{T}{|h|^{z / d}} \right)
\end{equation}
from which the thermodynamic quantities can be obtained. The scaling function $F(0)=$ constant to allow for a 
discontinuity in the magnetization at zero temperature as
the line of first order transitions is crossed.  The effects of couplings to dangerous irrelevant variables may be 
taken into account defining
$$
h = H-H_c-uT^{1/\psi_h}
$$
where in mean field $\psi_H = 1/2$ \cite{lawrie} or with another exponent to describe the $H_c$ dependence on
$(J/W)$Notice that this is the situation expected for a highly anisotropic, Ising-like material
\cite{dosSantos}.Notice that this is the situation expected for a highly anisotropic, Ising-like material
\cite{dosSantos}. Notice that the scaling expression, Eq.~\ref{free}, is useful for systems with a finite $T_N$
which are
driven to a  quantum critical point by the effect of an applied magnetic field. It does not apply at the quantum 
multicritical point at $H=0$.

We have pointed out important consequences of the application of Ehrenfest relation to antiferromagnetic materials
close to a quantum critical point. Generalized scaling relations were necessary to describe the system close to
the QCP where the N\'eel temperature vanishes. We have used some well known results from the phase diagram and
renormalization group studies of antiferromagnetic systems in a magnetic field to investigate the effects of this
field in the field-driven magnetic instability at zero temperature which may be realized, for example, in heavy 
fermion systems.

\begin{acknowledgments}
I would like to thank Eduardo Miranda and Enzo Granato for useful comments. This work is partially supported by
the Brazilian Agencies, FAPERJ and CNPq.
\end{acknowledgments}

\end{document}